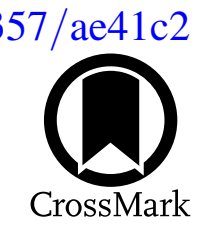

# Modeling Solitonic Cores, Stabilization of Bar, and Suppression of Bar Dissolution in DDO 168 via GPP Formalism: A Detailed Analysis of Bose–Einstein Condensate/Fuzzy Dark Matter Halo Structure and Bar Dynamics in the Dwarf Galaxy DDO 168

Saroj Khanal[1], Sanjay Kumar Sah[2], Kiran Khanal[3], and Sapana Khanal[1]
[1] Amrit Science Campus, Tribhuvan University, Kathmandu, Nepal
[2] Birendra Multiple Campus, Tribhuvan University, Chitwan, Nepal
[3] Patan Multiple Campus, Tribhuvan University, Kathmandu, Nepal

## Abstract

The cusp–core problem remains a challenge to the ΛCDM model, since dwarf galaxies often exhibit flat central density cores rather than the steep cusps ($\rho \propto r^{-1}$) predicted by collisionless *N*-body simulations. We model the dark-matter-dominated dwarf irregular galaxy DDO 168 within the Bose–Einstein condensate (BEC) or fuzzy dark matter (FDM) framework, in which ultralight bosons form a solitonic core governed by the Gross–Pitaevskii–Poisson (GPP) equations, with the soliton mass–radius relation enforced. We numerically validate the ground-state solution of the GPP system as a consistency check and fit the inner rotation curve of DDO 168 using SPARC data. Within this framework, the data are consistent with an axion mass $m = (1.3^{+0.3}_{-0.2}) \times 10^{-23}$ eV, and yield a solitonic core with characteristic radius $R_c = 2.40^{+0.22}_{-0.24}$ kpc, enclosing a mass $M(<2.47\ \text{kpc}) \simeq (1.5 \pm 0.2) \times 10^9 M_\odot$. The observed flat inner rotation curve is reproduced and the presence of a weak H I bar is compatible with multigigayear survival timescales, consistent with reduced Chandrasekhar dynamical friction in a shallow central potential. These results demonstrate that the BEC/FDM framework provides an internally consistent description of DDO 168, simultaneously reproducing the observed rotation curve, alleviating the cusp–core tension, and allowing long-lived weak bars under conservative dynamical assumptions.

## 1. Introduction

The standard ΛCDM model successfully describes large-scale structure but faces challenges at galactic scales, notably the cusp–core problem (W. J. G. de Blok 2010; J. S. Bullock & M. Boylan-Kolchin 2017). *N*-body simulations predict steep central density cusps ($\rho \propto r^{-1}$; J. F. Navarro et al. 1996), whereas dwarf and low-surface-brightness galaxies often exhibit flat cores (R. Kuzio de Naray & T. Kaufmann 2011; S.-H. Oh et al. 2011).

Several alternatives to cold dark matter (CDM) have been proposed, including self-interacting dark matter (D. N. Spergel & P. J. Steinhardt 2000), warm dark matter (P. Bode et al. 2001), and fuzzy or ultralight boson dark matter (W. Hu et al. 2000; L. Hui et al. 2017). In the latter scenario, ultralight bosons with masses of order $10^{-22}$ eV form Bose–Einstein condensates (BECs) supported by quantum pressure, giving rise to stable solitonic cores on kiloparsec scales that suppress cusp formation (D. J. E. Marsh & J. Silk 2014; H.-Y. Schive et al. 2014a, 2014b). These cores are governed by the Gross–Pitaevskii–Poisson (GPP) equations and naturally explain the observed flat inner rotation curves of dwarf galaxies (V. H. Robles & T. Matos 2012; E. Calabrese & D. N. Spergel 2016).

A substantial body of work has shown that solitonic or fuzzy dark matter (FDM) cores can reproduce inner rotation curves of dwarf and low-surface-brightness galaxies, typically using idealized halo models or statistical samples (e.g., V. H. Robles & T. Matos 2012; D. J. E. Marsh & J. Silk 2014; H.-Y. Schive et al. 2014a, 2014b). Subsequent studies have emphasized scaling relations between soliton core size, host halo mass, and boson mass, often focusing on ensemble behavior rather than individual systems (e.g., D. J. E. Marsh & A.-R. Pop 2015; L. Hui et al. 2017). In contrast, the present work adopts a galaxy-specific approach: we fit the official SPARC rotation-curve data of the weakly barred dwarf galaxy DDO 168 while explicitly enforcing the soliton mass–radius relation predicted by FDM simulations, and couple these kinematic constraints to an observation-driven analysis of bar dynamics. This combined treatment allows us to test whether a solitonic core can reproduce the observed rotation curve and remain dynamically compatible with the long-lived weak bar in the same system.

Galactic bars, common in disk galaxies, are sensitive to the inner halo profile. Cuspy halos induce strong dynamical friction (V. P. Debattista & J. A. Sellwood 2000), whereas cored halos prolong bar lifetimes (J. Shen & J. A. Sellwood 2004; A. Collier et al. 2019). The dwarf irregular galaxy DDO 168, a dark-matter-dominated system with a weak bar (N. N. Patra & C. J. Jog 2019), provides an ideal testbed. Its maximum rotation is ∼67 km s$^{-1}$; within the optical radius, the galaxy is strongly dark-matter dominated, with baryons contributing only (3–4) × $10^8 M_\odot$ to the total dynamical mass.

Here, we model DDO 168's central solitonic core via the GPP formalism, fit its rotation curve, and investigate bar stability. We show that a solitonic core can simultaneously reproduce the observed flat inner rotation curve and remain dynamically compatible with the weak bar observed in this system, reinforcing BEC dark matter as a plausible solution to small-scale ΛCDM tensions.

## 2. Theoretical Framework

### 2.1. Bose–Einstein Condensation

The main BEC dark matter candidates are ultralight scalar bosons (e.g., axions) with tiny masses and long de Broglie wavelengths ($\lambda_{\rm dB} \approx$ 2–4 kpc), forming coherent condensates on galactic scales (S. Das & R. K. Bhaduri 2014). Their gravitational effects create smooth, cored halo density profiles, matching observations in dwarf and low-mass galaxies better than standard CDM models.

### 2.2. GPP Formalism for DDO 168

The GPP formalism models ultralight bosonic dark matter by coupling the Gross–Pitaevskii equation for the condensate wave function $\psi(\boldsymbol{r}, t)$ with the Poisson equation for the gravitational potential (T. Rindler-Daller 2023). The Gross–Pitaevskii equation is

$$i\hbar\frac{\partial\psi}{\partial t} = \left[-\frac{\hbar^2}{2m}\nabla^2 + V_{\rm ext}(\boldsymbol{r}) + g\,|\psi|^2\right]\psi. \quad (1)$$

Here, $m$ is the mass of the bosons, $V_{\rm ext}(\boldsymbol{r})$ is the external potential (including the gravitational potential), and $g$ is the strength of the particle self-interaction. In the self-gravitating case, $V_{\rm ext}(\boldsymbol{r})$ is replaced by the gravitational potential $\Phi(\boldsymbol{r})$.

The gravitational potential is determined self-consistently through the Poisson equation,

$$\nabla^2\Phi = 4\pi G\rho, \quad (2)$$

where $\rho = m|\psi|^2$ is the density of the dark matter condensate and $G$ is the gravitational constant.

### 2.3. Numerical Validation of the Solitonic Core Profile

Although the solitonic core profile is constrained directly from the SPARC rotation-curve data using the analytic soliton profile, we also solve the time-independent GPP equations to verify that this profile accurately represents the ground-state solution. This numerical calculation serves as a consistency check rather than an independent fitting procedure.

The spherically symmetric GPP system is solved on a one-dimensional radial grid extending to $R_{\rm max} = 16$ kpc with uniform spacing $\Delta r \simeq 0.02$ kpc. Regularity at the origin and vanishing boundary conditions at large radii are imposed. Doubling the spatial resolution changes the density and enclosed-mass profiles by less than ∼2% within the soliton-dominated region, confirming numerical convergence. The resulting ground-state density profile is well reproduced by the analytic soliton form adopted in the rotation-curve analysis.

## 3. Data and Methodology

We model the dark matter halo of DDO 168 as a ground-state solitonic core within the BEC/FDM framework. We adopt the standard analytic soliton profile to fit the SPARC rotation curve. The gravitational potential is treated in the Newtonian regime. Although baryonic contributions are subdominant over the fitted radial range (0.4–2.47 kpc), we assess the sensitivity of the inferred solitonic parameters to plausible baryonic uncertainties. For DDO 168, the combined baryonic contribution to the circular velocity remains below ≲15% across the fitted region (F. Lelli et al. 2016; N. N. Patra & C. J. Jog 2019). Varying the stellar mass-to-light ratio and gas normalization within conservative ranges (±30%) produces changes in the inferred soliton core radius and axion mass that remain within the statistical uncertainties inferred from the MCMC analysis, once parameter degeneracies and sampling variance are taken into account. Consequently, baryonic uncertainties do not dominate the soliton parameter inference for DDO 168.

**Table 1**
Official SPARC Rotation-curve Measurements for the Dwarf Galaxy DDO 168 (F. Lelli et al. 2016) Used to Constrain the Solitonic Core Model

| Radius (kpc) | $V_{\rm obs}$ (km s$^{-1}$) | $\sigma_{V_{\rm obs}}$ (km s$^{-1}$) |
|---|---|---|
| 0.41 | 15.00 | 4.10 |
| 0.82 | 23.40 | 1.30 |
| 1.24 | 29.20 | 1.50 |
| 1.64 | 36.20 | 1.10 |
| 2.06 | 44.00 | 1.10 |
| 2.47 | 52.10 | 1.20 |

**Note.** Listed are the galactocentric radius $R$, the observed circular velocity $V_{\rm obs}$, and the corresponding SPARC uncertainty $\sigma_{V_{\rm obs}}$. Only data points within $R \leqslant 2.47$ kpc are shown, corresponding to the dark-matter-dominated region that directly constrains the solitonic core parameters.

### 3.1. SPARC Rotation-curve Data

We use the official rotation-curve data for DDO 168 from the SPARC database (F. Lelli et al. 2016), which provides rotation velocities derived from high-quality H I velocity fields together with formally estimated observational uncertainties. The SPARC rotation curve of DDO 168 extends to an outermost measured radius of $R = 2.47$ kpc, fully covering the region where the galaxy is strongly dark-matter dominated and where the solitonic core is expected to govern the gravitational potential.

For the present analysis, we restrict the fit to the inner region $0.4 \lesssim R \lesssim 2.47$ kpc, which directly constrains the solitonic core. Within this framework, the rotation-curve data constrain the axion mass through its impact on the soliton scale. Table 1 lists the subset of official SPARC data points used in the fitting procedure, including the quoted uncertainties in the rotation velocity.

The observed circular velocities trace the enclosed dynamical mass via

$$M(<R) = \frac{R\,V_{\rm obs}^2}{G}, \quad (3)$$

providing a direct, model-independent constraint on the mass distribution within the fitted radial range.

### 3.2. Error Model

The SPARC database reports formal uncertainties that account for measurement noise, beam smearing, inclination uncertainties, and noncircular motions. In dwarf irregular galaxies such as DDO 168, additional systematic effects can contribute at the level of a few km s$^{-1}$.

To account for these effects conservatively, we adopt the SPARC-reported uncertainties and include a fixed additional systematic term added in quadrature when evaluating the goodness of fit, following standard practice in dwarf-galaxy rotation-curve analyses. This approach avoids artificially small uncertainties while remaining consistent with standard treatments of dwarf galaxy rotation curves (e.g., J. A. Sellwood & S. S. McGaugh 2005). With this error model, reduced chi-square values $\chi^2_\nu \lesssim 1$ indicate statistically acceptable fits. Our conclusions are unchanged for reasonable variations of this systematic term within the range 1–3 km s$^{-1}$.

### 3.3. Goodness-of-fit Statistics

The goodness of fit is quantified using the chi-square statistic,

$$\chi^2 = \sum_{i=1}^{N} \frac{[V_{\rm obs}(R_i) - V_{\rm model}(R_i)]^2}{\sigma_i^2}, \tag{4}$$

where $V_{\rm obs}(R_i)$ and $V_{\rm model}(R_i)$ are the observed and model circular velocities at radius $R_i$, respectively, and $\sigma_i$ includes the SPARC-reported observational uncertainty with an additional systematic term added in quadrature, as described in Section 3.2.

The reduced chi-square is defined as

$$\chi^2_\nu = \frac{\chi^2}{\nu}, \tag{5}$$

where $\nu = N - N_{\rm par}$ is the number of degrees of freedom.

For DDO 168, $N = 6$ rotation-curve data points are fitted using two correlated solitonic parameters (the axion mass $m$ and soliton core radius $R_c$) under the enforced soliton scaling relation, yielding $\nu = 4$ degrees of freedom.

### 3.4. Solitonic Density Profile and Model Construction

The dark matter halo of DDO 168 is modeled using the solitonic density profile predicted by FDM simulations and by solutions of the GPP equations. We adopt the analytic soliton profile

$$\rho(r) = \frac{\rho_0}{\left[1 + 0.091\left(\frac{r}{R_c}\right)^2\right]^8}, \tag{6}$$

with the theoretical scaling

$$\rho_0 = 1.9 \times 10^9 \left(\frac{m}{10^{-23}\,{\rm eV}}\right)^{-2} \left(\frac{R_c}{\rm kpc}\right)^{-4} M_\odot\ {\rm kpc}^{-3}, \tag{7}$$

from (H.-Y. Schive et al. 2014a), enforcing this relation during the MCMC fitting to constrain the boson mass $m$. This relation enforces a physical scaling between the soliton parameters, reducing the freedom in the model but still leaving two correlated fit parameters: the axion mass $m$ and the core radius $R_c$.

The enclosed-mass profile is obtained via numerical integration,

$$M(<r) = \int_0^r 4\pi r'^2\, \rho(r')\, {\rm d}r', \tag{8}$$

and the corresponding circular velocity is

$$V_{\rm sol}(r) = \sqrt{\frac{G\, M(<r)}{r}}. \tag{9}$$

Within the radial range probed by the SPARC data, the solitonic core provides an adequate description of the gravitational potential under the enforced soliton scaling relation. At larger radii, beyond the data range used for fitting, the soliton is not expected to represent the full halo structure.

### 3.5. Parameter Estimation

We constrain the solitonic core through the axion mass $m$ and the soliton core radius $R_c$, using both direct $\chi^2$ minimization and a Bayesian Markov Chain Monte Carlo (MCMC) analysis. The likelihood function is constructed from the difference between the observed SPARC rotation velocities and the corresponding model predictions, adopting the error model described above. The MCMC analysis yields well-behaved, unimodal posterior distributions after increasing the total number of MCMC steps to ensure adequate sampling, and reveals a clear degeneracy between $m$ and $R_c$, reflecting the intrinsic scaling relation of solitonic solutions in the FDM framework. Despite this degeneracy, the resulting rotation curve and enclosed-mass profile remain tightly constrained within the observed radial range. In the Bayesian analysis, we adopt a log-uniform prior on the axion mass over $10^{-23} \leqslant m \leqslant 10^{-21}$ eV, motivated by previous studies of FDM in dwarf galaxies (e.g., H.-Y. Schive et al. 2014a; D. J. E. Marsh & A.-R. Pop 2015; L. Hui et al. 2017), and a uniform prior on the soliton core radius over $0.5 \leqslant R_c \leqslant 5.0$ kpc, encompassing the radial range probed by the SPARC data. The lower prior bound $m \geqslant 10^{-23}$ eV is physically motivated by the radial extent of the rotation-curve data. For smaller boson masses, the soliton mass–radius scaling implies core radii larger than the outermost measured point ($R = 2.47$ kpc), rendering the solitonic profile effectively unconstrained by the data and producing an approximately solid-body rotation curve over the fitted region. Such behavior is inconsistent with the observed slope of the inner rotation curve. Extending the prior to lower masses therefore adds only unconstrained parameter volume without altering the inferred enclosed-mass profile or the quality of the fit.

### 3.6. Linking Solitonic Core to Bar Dynamics

The solitonic core inferred from the rotation-curve analysis modifies the inner gravitational potential of DDO 168 by producing a shallow central density profile relative to cuspy CDM halos. Such cored inner potentials are known to reduce angular-momentum transfer between stellar bars and dark matter halos, thereby suppressing rapid bar slowdown and dissolution.

In the following section, we use the axion-mass–constrained solitonic density profile to estimate characteristic dynamical-friction timescales for the weak bar in DDO 168 and to assess whether the inferred inner halo structure is consistent with the observed longevity of the bar.

## 4. Results and Discussion

Using the solitonic core model described in Section 3, with the soliton scaling relation enforced, we fit the SPARC rotation curve of DDO 168 and obtain a statistically acceptable description of the observed inner kinematics. Figure 1 shows the best-fitting solitonic rotation curve together with the SPARC data points and residuals.

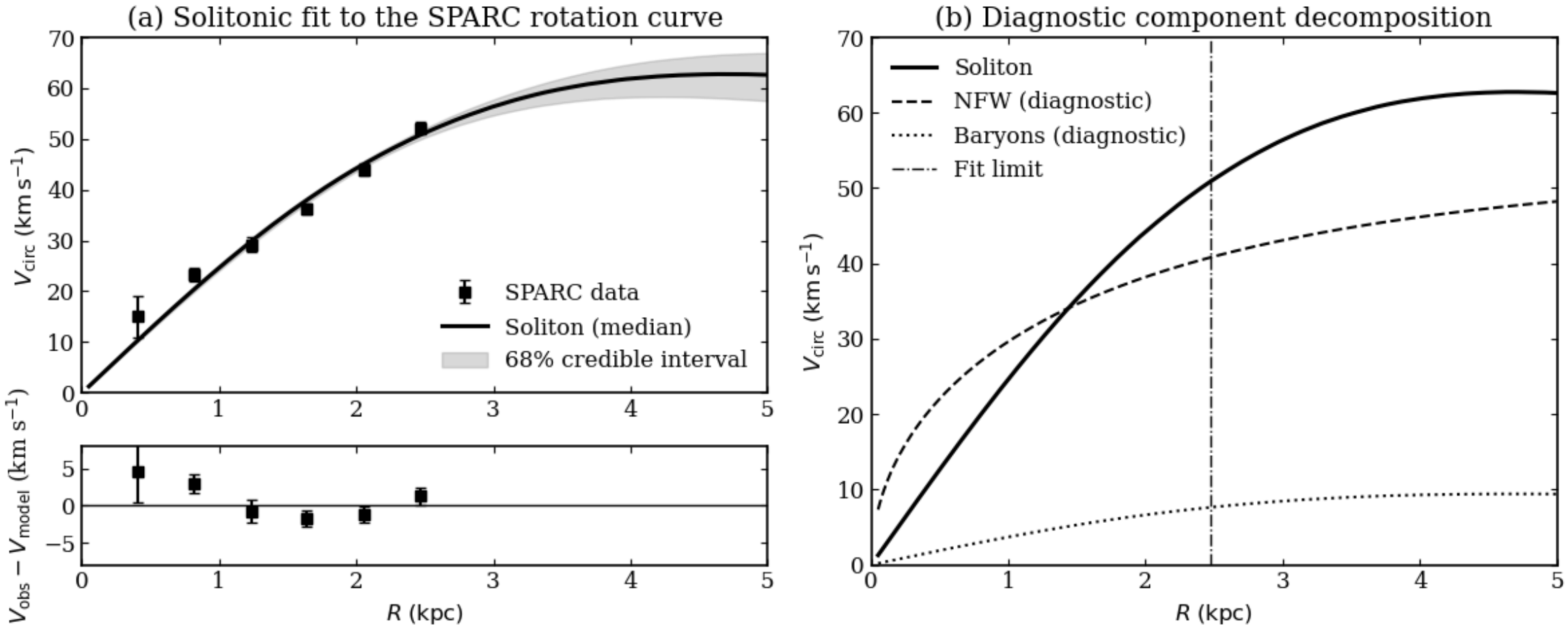

**Figure 1.** Panel (a): Solitonic fit to the SPARC rotation curve of DDO 168. Black points show the observed circular velocities with $1\sigma$ uncertainties. The solid line shows the median posterior solitonic model, and the shaded gray region indicates the 16%–84% posterior predictive interval derived from the MCMC samples. The lower panel shows residuals ($V_{\rm obs} - V_{\rm model}$). Panel (b): Diagnostic decomposition of the circular velocity into the solitonic dark-matter component (solid line), an illustrative NFW halo (dashed line), the baryonic contribution (dotted line), and the radial fit limit (vertical dash–dotted line at $R = 2.47$ kpc).

### *4.1. DDO 168 Rotation Curve with BEC Soliton*

Figure 1(a) demonstrates that the solitonic core model reproduces both the inner rise and the subsequent gradual flattening of the rotation curve of DDO 168 over the radial range probed by the SPARC data ($0.4 \lesssim R \lesssim 2.47$ kpc), without invoking additional halo components. For clarity, the best-fitting solitonic model is shown out to $R \approx 2.5$ kpc and corresponds to a soliton core radius $R_c \simeq 2.4$–$2.6$ kpc, consistent with the median and credible interval of the MCMC posterior. The residuals remain small and scatter symmetrically about zero, with no evidence for a systematic radial trend. The rms scatter is approximately 2–3 km s$^{-1}$, well within the quoted SPARC uncertainties, yielding reduced chi-square values $\chi^2_\nu \lesssim 1$. The absence of systematic residuals is consistent with the presence of a finite-density core in the inner halo of DDO 168, as expected in FDM and BEC models.

### *4.2. Diagnostic Testing of a Composite Soliton + NFW + Baryon Model for DDO 168*

Because the gravitational potential obeys a linear Poisson equation, the circular-velocity contributions of the solitonic core, outer halo, and baryonic disk can be expressed as a formal sum,

$$V_{\rm tot}^2(r) = V_{\rm sol}^2(r) + V_{\rm NFW}^2(r) + V_{\rm bar}^2(r). \quad (10)$$

To extend the solitonic-core analysis beyond the inner region probed by the SPARC data and assess the role of additional mass components at larger radii, we consider a composite mass model consisting of three elements:

1. A quantum-pressure-supported solitonic core,
2. An outer NFW-like dark matter halo, and
3. The observed baryonic disk (stars + gas).

The emphasis here is on illustrating the relative radial contributions of each component rather than on constructing a global best-fit rotation curve.

The solitonic core parameters are fixed to the Bayesian median values inferred from the inner rotation-curve analysis (Section 4.3), ensuring consistency with the soliton-dominated region directly constrained by the SPARC data, with the median values shown for illustration; using other posterior samples produces indistinguishable curves within the fitted region. The NFW scale density and scale radius are adopted as $\rho_s = 6.91 \times 10^6 M_\odot$ kpc$^{-3}$ and $r_s = 5.82$ kpc from the best-fit dark matter halo model of DDO 168 in the SPARC database (F. Lelli et al. 2016; H. Katz et al. 2017). These NFW parameters are adopted without refitting and are used solely to illustrate the outer-halo contribution beyond the region dominated by the solitonic core. The baryonic disk contribution, as quantified in Section 3, contributes less than ~15% of the total circular velocity across the fitted region.

Figure 1 (b) shows the individual contributions of the solitonic core, NFW halo, and baryonic components to the rotation curve of DDO 168 using the full SPARC dataset. The solitonic core reproduces the observed inner rotation curve within its fitted domain ($R \leqslant 2.47$ kpc), while the NFW halo becomes increasingly important at larger radii.

### *4.3. Parameter Uncertainties*

#### *4.3.1. Parameter Uncertainties from MCMC Analysis*

A Bayesian MCMC analysis was performed to obtain posterior constraints on the solitonic core parameters under the enforced soliton mass–radius scaling relation. The marginalized posterior distributions yield an axion mass

$$\log_{10}\left(\frac{m}{\rm eV}\right) = -22.90^{+0.08}_{-0.07}, \quad (11)$$

and a corresponding soliton core radius

$$R_c = 2.40^{+0.22}_{-0.24}\ {\rm kpc}, \quad (12)$$

where the quoted uncertainties represent the 16%–84% credible intervals. The posterior is prior-limited on the low-mass side because the available kinematic data do not extend far enough to constrain soliton cores larger than ~3 kpc.

We emphasize that the moderate width of the posterior distributions primarily reflects enforcement of the soliton mass–radius scaling relation intrinsic to the FDM framework, rather than independent constraining power from the limited number of inner rotation-curve data points. The strong anticorrelation between $m$ and $R_c$ seen in the joint posterior is a direct consequence of this physical scaling. Additional systematic uncertainties in distance, inclination, and noncircular motions may further broaden the true uncertainties but are not expected to qualitatively alter the inferred inner mass distribution.

The Bayesian posterior exhibits a well-defined high-probability region, within which the minimum-$\chi^2$ solution lies inside the 68% credible region. Because the soliton central density is determined self-consistently from $m$ and $R_c$, the Bayesian median and best-fit solutions differ only slightly and yield nearly identical enclosed-mass profiles within the observed radial range.

At the outermost measured radius $R = 2.47$ kpc, the enclosed dynamical mass is

$$M(<R) \simeq (1.5\text{–}1.6) \times 10^9\, M_\odot, \tag{13}$$

with variations at the $\lesssim$10% level across representative posterior samples. Within the radial range directly constrained by the SPARC data, the MCMC fit reproduces the observed rotation curve with high fidelity, indicating that the inferred gravitational potential is robust despite intrinsic parameter correlations.

#### 4.3.2. Conclusion with MCMC

The resulting posterior distributions and parameter correlations obtained from the Bayesian MCMC analysis are displayed in Figure 2.

The Bayesian median and minimum-$\chi^2$ solitonic models shown in the figure are derived from the same MCMC sampling, and both provide acceptable fits to the SPARC rotation curve. The posterior widths visible in Figure 2 reflect the enforced soliton scaling relation and the limited number of rotation-curve data points, rather than unusually strong independent constraining power. Despite this intrinsic degeneracy, physically relevant quantities such as the enclosed-mass profile remain robust across the posterior.

### 4.4. Bar Stabilization

A solitonic core is expected to reduce bar–halo dynamical friction and thereby favor bar survival over gigayear timescales. We investigate bar stabilization using observational constraints from the SPARC database (F. Lelli et al. 2016), and these observational constraints are found to be consistent with the solitonic core parameters inferred from the rotation-curve and MCMC analysis, using posterior median values for the solitonic core. The adopted parameters are as follows:

1. Bar mass:
$$M_{\rm bar} \approx 10^7\, M_\odot, \tag{14}$$
which is indirectly estimated and allowed to vary by $\pm$50% (S.-H. Oh et al. 2015; N. N. Patra & C. J. Jog 2019).
2. Bar radius:
$$R_{\rm bar} \approx 1.0\ \text{kpc}, \tag{15}$$
with circular velocity $V \approx 25$–$30\ \text{km s}^{-1}$ inferred from the rotation curve at this radius, corresponding to a characteristic pattern speed consistent with these circular velocities (F. Lelli et al. 2016).
3. Local dark matter density at $R_{\rm bar} \approx 1.0$ kpc:
$$\rho_h \approx (3\text{–}5) \times 10^6\, M_\odot\ \text{kpc}^{-3}, \tag{16}$$
where $\rho_h$ denotes the effective background density entering the Chandrasekhar dynamical friction formalism. This value is derived from the best-fitting solitonic core profile, defined by the correlated parameters $m$ and $R_c$, evaluated at the bar radius and reduced relative to the total soliton density to account for shallow phase-space gradients and suppressed wake formation characteristic of cored halos (L. Hui et al. 2017).
4. Halo velocity dispersion:
$$\sigma_h \approx 15\text{–}25\ \text{km s}^{-1}, \tag{17}$$
adopted as a fiducial range (J. A. Sellwood & S. S. McGaugh 2005).
5. Coulomb logarithm:
$$\ln\Lambda \sim 3\text{–}5 \tag{18}$$
(S. Tremaine & M. D. Weinberg 1984).

The soliton core is several times denser than the effective background halo at the bar radius,

$$\frac{\rho_{\rm sol}(R \simeq R_{\rm bar})}{\rho_h} \sim 7\text{–}12, \tag{19}$$

indicating that the soliton core dominates the inner gravitational potential, while the background halo contributes primarily at larger radii (L. Hui et al. 2017).

#### 4.4.1. Estimation of the Bar Mass Parameter

To evaluate the impact of the solitonic core on bar dynamics in DDO 168, we adopt Chandrasekhar's dynamical friction formalism as a first-order, order-of-magnitude estimate, treating the bar as an effective point mass. This approximation is appropriate for a weak, slowly rotating gaseous bar. While originally derived for compact perturbers in homogeneous backgrounds, this formalism is commonly employed as a qualitative diagnostic of angular-momentum exchange in systems hosting weak gaseous bars, where nonaxisymmetric perturbations are relatively mild (V. P. Debattista & J. A. Sellwood 2000; A. Collier et al. 2019).

The bar mass,

$$M_{\rm bar} \sim 10^7\, M_\odot, \tag{20}$$

is estimated from H I observations by integrating the neutral gas surface density over the bar region using H I column-density maps from the Very Large Array LITTLE-THINGS survey (S.-H. Oh et al. 2015; N. N. Patra & C. J. Jog 2019). Given the gas-dominated nature of DDO 168 and the absence of a significant stellar bar, the H I mass provides the dominant contribution to the bar mass.

The bar mass is estimated as

$$M_{\rm bar} = \int \Sigma_{\rm H\,I}(r, \theta)\, dA \approx \langle \Sigma_{\rm H\,I} \rangle A_{\rm bar}, \tag{21}$$

where $A_{\rm bar}$ is the deprojected bar area and $\langle \Sigma_{\rm H\,I} \rangle$ is the mean H I surface density.

Uncertainties in the mean H I surface density and bar geometry, arising from inclination ($\sim$66°) and beam-smearing effects, correspond to $\sim$20%–30% variations in $\langle \Sigma_{\rm H\,I} \rangle$ and $A_{\rm bar}$. These propagate linearly into $M_{\rm bar}$ and are fully encompassed by the conservative ±50% mass range adopted here. The bar feature has a projected length of ∼1.9 kpc and a width of ∼0.5 kpc, corresponding to a deprojected area of ∼0.95 kpc$^2$ for an adopted bar radius $R_{\mathrm{bar}} \approx 1$ kpc.

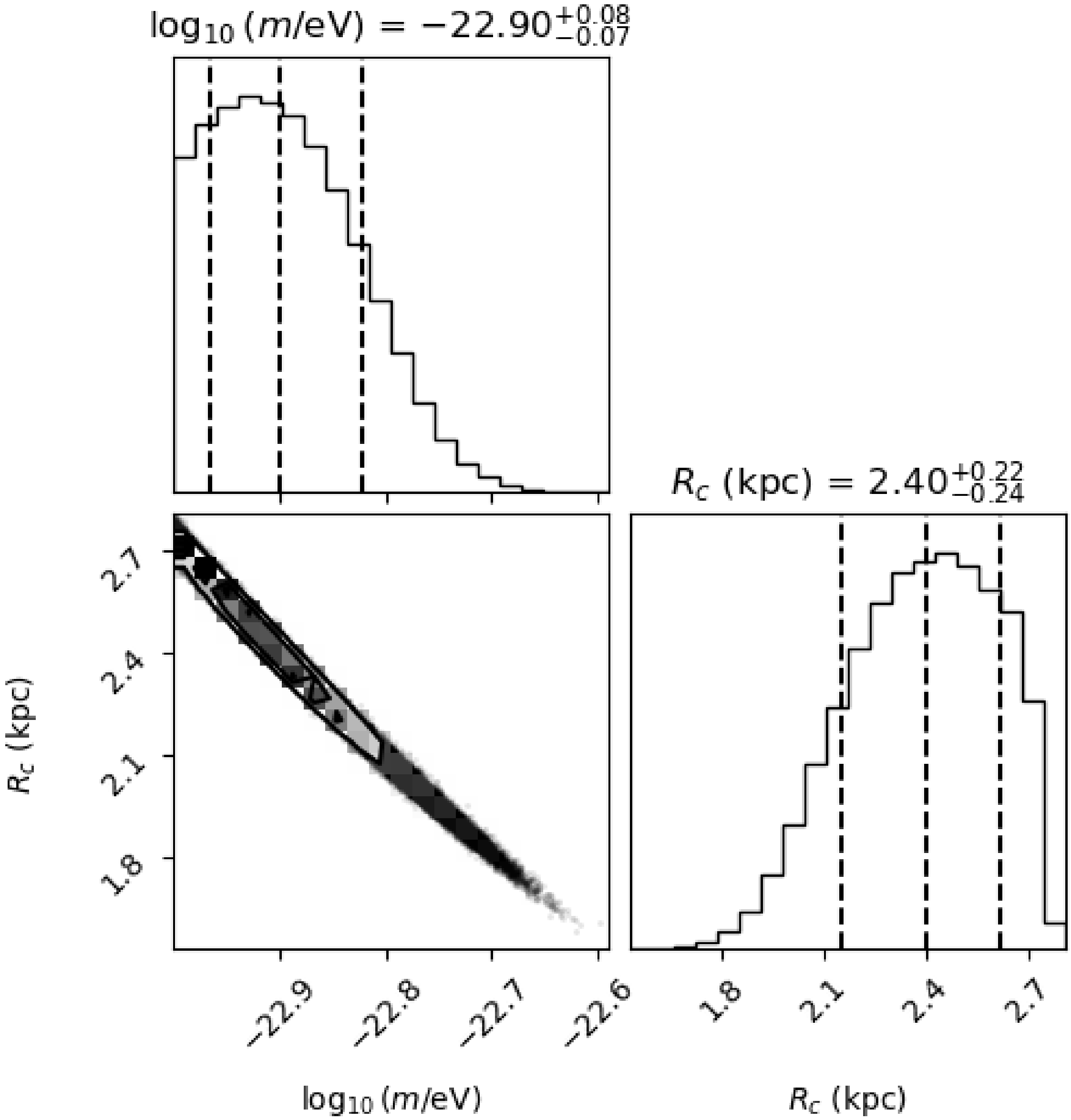

**Figure 2.** Posterior distributions and parameter correlations for DDO 168 from the Bayesian MCMC analysis. The apparent truncation of the posterior at low axion mass reflects the physically motivated prior bound $m \geqslant 10^{-23}$ eV; lower masses correspond to soliton cores exceeding the radial extent of the data and are therefore unconstrained by the rotation-curve measurements. Top-left panel: Marginalized posterior distribution for the axion mass $\log_{10}(m/\mathrm{eV})$, with vertical lines marking the Bayesian median value and the 16%–84% credible interval. Bottom-left panel: Marginalized posterior distribution for the solitonic core radius $R_c$, with vertical lines marking the Bayesian median value and the 16%–84% credible interval. Bottom-right panel: Joint posterior samples exhibiting a strong correlation between the axion mass and the solitonic core radius, reflecting the intrinsic mass–scaling degeneracy of solitonic dark matter cores.

H I column-density maps indicate characteristic surface densities in the bar region of order

$$\Sigma_{\mathrm{H\,I}} \sim 10\, M_\odot\ \mathrm{pc}^{-2} \approx 10^7\, M_\odot\ \mathrm{kpc}^{-2}, \tag{22}$$

after accounting for helium (P. M. W. Kalberla & J. Kerp 2009), consistent with the mean surface density adopted in the bar mass estimate. The inferred bar mass is

$$M_{\mathrm{bar}} \sim (0.9\text{–}1.4) \times 10^7\, M_\odot, \tag{23}$$

corresponding to ∼3%–5% of the total H I mass of the galaxy (∼2.8 × $10^8 M_\odot$; S.-H. Oh et al. 2015), consistent with the observed weak bar strength $s \approx 0.2$ (N. N. Patra & C. J. Jog 2019).

To account for uncertainties associated with the extended nature of the bar and the simplified friction formalism, we vary $M_{\mathrm{bar}}$ by ±50% around the fiducial value. Within this range, the reduced dynamical friction expected in a cored halo derived from our soliton density profile yields friction timescales consistent with long-lived bars, in qualitative agreement with numerical studies of bar evolution in cored dark matter halos (J. Shen & J. A. Sellwood 2004; A. Collier et al. 2019).

#### 4.4.2. Bar Stabilization Using Chandrasekhar Dynamical Friction Estimate (for Bar in Halo)

Although solitonic cores are expected to reduce dynamical friction relative to cuspy halos, a residual frictional force can persist when a massive perturber such as a galactic bar moves through the surrounding dark matter halo (R. Boey et al. 2024; K. Blum et al. 2025). To assess the solitonic core's influence on bar stability, the Chandrasekhar dynamical friction formalism is applied to estimate the dynamical-friction force and the associated torque acting on the bar at its characteristic radius ($R_{\rm bar} = 1.0$ kpc). This first-order approach enables an order-of-magnitude assessment of bar–halo dynamical friction within different inner halo profiles. The Chandrasekhar formalism is applied here as an approximate description for an extended bar evolving within a nonuniform dark-matter background. The Chandrasekhar formula for a perturber of mass $M_{\rm bar}$ moving at velocity $V$ through a background of effective density $\rho_h$ and velocity dispersion $\sigma_h$ is

$$F_{\rm df} = -4\pi G^2 M_{\rm bar}^2\, \rho_h\, \frac{\ln\Lambda}{V^2}\left[{\rm erf}(X) - \frac{2X}{\sqrt{\pi}}\, e^{-X^2}\right], \quad (24)$$

where the dimensionless parameter $X$ is defined as

$$X = \frac{V}{\sqrt{2}\,\sigma_h}. \quad (25)$$

Here, $F_{\rm df}$ is the dynamical-friction force, $G$ is the gravitational constant, and $M_{\rm bar}$ is the bar mass. The quantity $\rho_h$ is the effective background dark-matter density entering the Chandrasekhar formalism, appropriate for dynamical friction in a cored halo. It is evaluated at the bar radius from the solitonic core profile and reduced relative to the total soliton density, reflecting the shallow phase-space gradients and suppressed gravitational wake formation characteristic of cored and FDM halos (L. Hui et al. 2017; P. Mocz et al. 2017). This effective density is used in all bar-slowdown calculations at the bar radius $R_{\rm bar} = 1.0$ kpc.

$V$ is the characteristic relative velocity at $R_{\rm bar}$ (approximated by the local circular velocity) and $\ln\Lambda$ is the Coulomb logarithm characterizing the range of gravitational encounters. The term within brackets,

$$\left[{\rm erf}(X) - \frac{2X}{\sqrt{\pi}}\, e^{-X^2}\right], \quad (26)$$

is velocity-dependent.

The torque ($T$) exerted on the bar is estimated as

$$T \simeq R_{\rm bar}\, F_{\rm df}. \quad (27)$$

The corresponding bar-slowdown timescale is given by

$$\tau_{\rm bar} = \frac{L_{\rm bar}}{T}, \quad (28)$$

where the bar angular momentum is approximated as

$$L_{\rm bar} \simeq M_{\rm bar}\, R_{\rm bar}\, V. \quad (29)$$

An approximate characteristic slowdown timescale can then be written as

$$\tau_{\rm slow} = \frac{V^3}{4\pi G^2 M_{\rm bar}\, \rho_h\, \ln\Lambda\left[{\rm erf}(X) - \frac{2X}{\sqrt{\pi}}\, e^{-X^2}\right]}. \quad (30)$$

**Table 2**
Estimated Bar-slowdown Timescales in DDO 168 from Chandrasekhar-type Dynamical Friction

| $\ln\Lambda$ | $\sigma_{\rm h}$ (km s$^{-1}$) | $\tau$ (Gyr) |
|---|---|---|
| 3 | 15 | 2.1 |
| 3 | 20 | 5.0 |
| 3 | 30 | 16.8 |
| 2 | 20 | 7.5 |
| 5 | 20 | 3.0 |

**Note.** The quoted timescales span the full ±50% uncertainty range in the bar mass, which dominates over other observational uncertainties such as beam smearing and inclination.

For a galactic bar interacting with its dark matter halo, established scaling arguments suggest a Coulomb logarithm typically falling within the range $\ln\Lambda \sim 2$–4 (S. Tremaine & M. D. Weinberg 1984; M. D. Weinberg 1985). To explore the sensitivity of our results to this parameter, we consider the broader range $\ln\Lambda = 2$–5 in our analysis.

The solitonic core in DDO 168, characterized by a shallow density profile, reduces the effective phase-space density (Rindler-Daller 2023). A further stabilizing mechanism arises qualitatively from the quantum nature of the solitonic core. As demonstrated by the simulations of P. Mocz et al. (2017), wave-like density fluctuations may inject kinetic energy into the bar that counteracts frictional losses. The solitonic core thus mitigates angular-momentum loss, thereby favoring bar persistence.

We use representative posterior samples of the correlated parameters ($m$, $R_c$) together with plausible halo velocity dispersions to estimate the slowdown timescale $\tau$ at the order-of-magnitude level. Varying ($m$, $R_c$) across the 68% credible region produces changes in $\tau$ well within an order of magnitude, and does not alter the qualitative conclusion regarding bar survival. The obtained values of $\tau$ are consistent with the presence of a long-lived weak bar. This approach yields timescales qualitatively comparable to classical studies while explicitly capturing the density structure of the BEC/FDM soliton.

Although the soliton is a coherent quantum structure, residual particle-like interactions occur due to its granular nature on small scales (P. Mocz et al. 2017; T. Rindler-Daller 2023). Using a much lower diffuse halo envelope density, $\rho \simeq 4.64 \times 10^6 M_\odot\,{\rm kpc}^{-3}$ (F. Lelli et al. 2016), would underestimate dynamical friction and overestimate bar stability. Our approach instead adopts an effective background density $\rho_h$ together with a multiplicative correction applied to the Chandrasekhar point-mass slowdown timescale to account for the extended spatial structure of the bar and the reduced coherence of the induced gravitational wake. We adopt a conservative correction factor of order 10 to the point-mass Chandrasekhar slowdown timescale, consistent with the range of factors (≈5–15) used in analytical and numerical studies of extended bars in cored halos (M. D. Weinberg 1985; V. P. Debattista & J. A. Sellwood 2000).

Using the Chandrasekhar formalism, we estimate the bar-slowdown timescale as presented in Table 2. The Chandrasekhar formulation describes dynamical friction for a point-mass perturber moving through a background medium. However, for an extended structure such as a galactic bar, the effective frictional torque is reduced because the induced gravitational wake is more diffuse. Following the standard approach in the literature (M. D. Weinberg 1985; V. P. Debattista & J. A. Sellwood 2000), we adopt a scaling factor of 10 applied to the point-mass dynamical-friction timescale to account for the bar's finite spatial extent and the diminished wake response in a cored halo potential. This correction is conservative in analytical estimates of bar–halo dynamical friction.

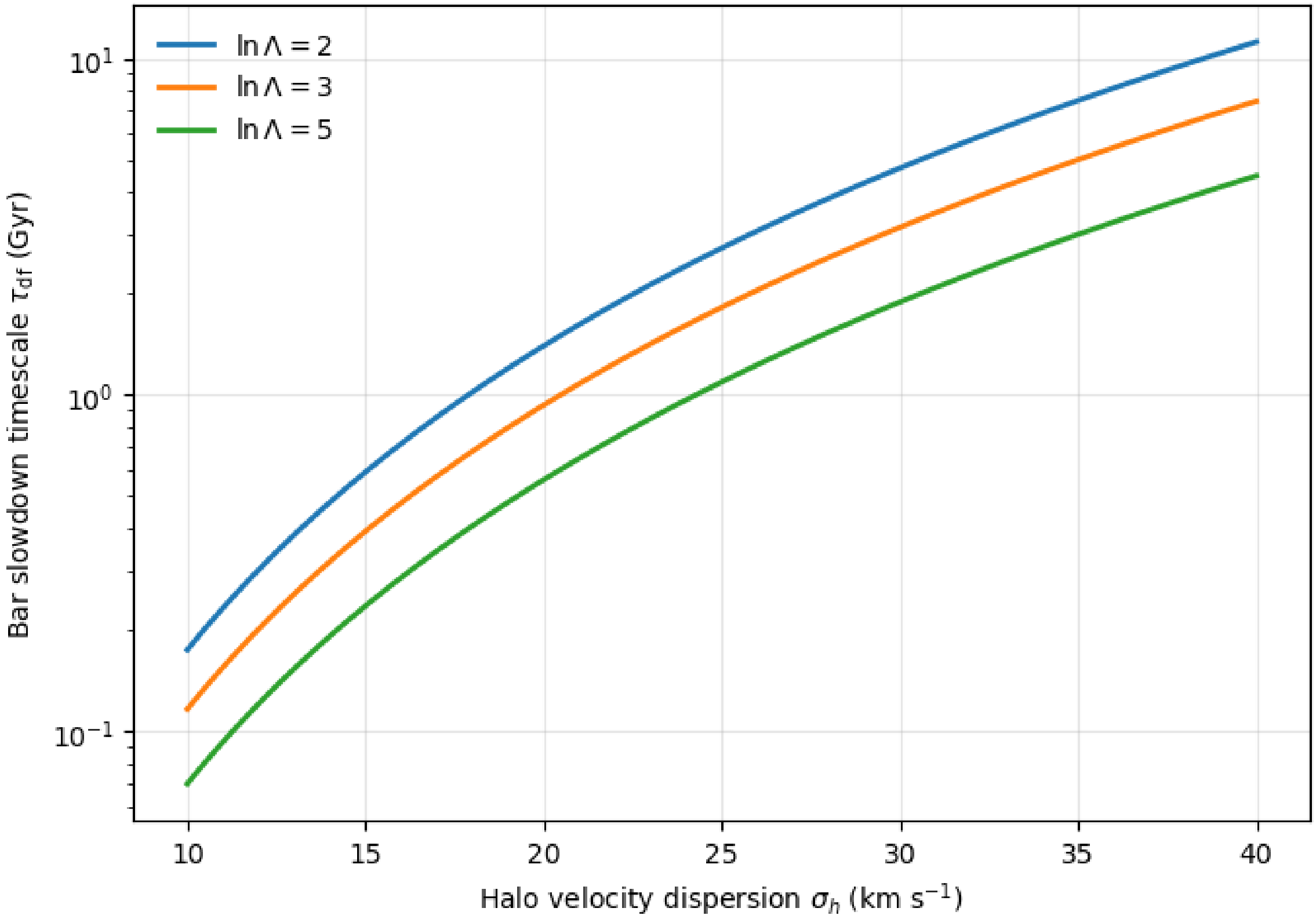

**Figure 3.** Bar-slowdown timescale $\tau$ as a function of the Coulomb logarithm ln $\Lambda$ for different values of $\sigma_h$ (order-of-magnitude estimates).

In Table 2, the halo velocity dispersion $\sigma_h$ is used to parameterize the characteristic relative velocity scale of the bar–halo interaction, such that the Chandrasekhar slowdown timescale scales as $\tau_{\rm df} \propto \sigma_h^3$ for fixed halo density and Coulomb logarithm.

### 4.4.3. Explanation of Table 2

For the fiducial choice $\ln\Lambda = 3$ (a conservative value) and $\sigma_h = 20\,{\rm km\,s^{-1}}$, the bar-slowdown timescale is $\tau \approx 5$ Gyr. This corresponds to a characteristic slowdown timescale of order ∼5 Gyr, assuming approximately steady dynamical friction, implying a reduction of the bar pattern speed to ∼37% of its initial value over this timescale, and to ∼14% after ∼10 Gyr. This behavior is consistent with the long-term structural coherence of the bar, in agreement with the survival trends shown in Figure 3.

The dependence of the slowdown timescale $\tau$ on the halo velocity dispersion $\sigma_h$ and the Coulomb logarithm $\ln\Lambda$ is illustrated in Figure 4. For fixed $\ln\Lambda$, $\tau$ increases steeply with $\sigma_h$, while for fixed $\sigma_h$, larger values of $\ln\Lambda$ lead to systematically shorter slowdown timescales.

Accordingly, the Chandrasekhar-based slowdown timescales presented here should be interpreted as approximate estimates for an extended bar embedded in a nonuniform, soliton-dominated halo.

### 4.4.4. Suppression of Bar Dissolution

The solitonic core in DDO 168 suppresses bar dissolution by extending the characteristic bar-slowdown timescale well into the multigigayear regime. At the characteristic bar radius $R_{\rm bar} \simeq 1.0$ kpc, the inferred effective background density is $\rho_{\rm h} \sim (3–5) \times 10^6 M_\odot\,{\rm kpc}^{-3}$. Conservative corrections accounting for the extended nature of the bar are included following M. D. Weinberg (1985) and V. P. Debattista & J. A. Sellwood (2000). In addition to the extended-bar correction, dissolution-efficiency scalings motivated by numerical simulations (e.g., A. Collier et al. 2019) are adopted qualitatively to account for the fact that bar dissolution proceeds more slowly than angular-momentum loss alone.

Even under pessimistic parameter choices, such as large Coulomb logarithms ($\ln\Lambda \simeq 5$) and maximal bar-mass assumptions, the resulting bar-slowdown timescale remains several gigayears, allowing the bar to survive for multigigayear timescales. In contrast, equivalent calculations for a cuspy Navarro–Frenk–White (NFW) halo yield rapid bar dissolution on subgigayear timescales. This quantitative comparison shows that the quantum-pressure-supported solitonic core not only reproduces the observed inner rotation curve of DDO 168 but also mitigates dynamical-friction-driven bar dissolution. The resulting suppression of bar slowdown and enhanced bar survival are illustrated in Figure 4.

Figure 3 emphasizes parameter sensitivity, whereas Figure 4 illustrates the corresponding qualitative difference between solitonic and cuspy halos. Crucially, our conclusions do not rely on the precise numerical value of the slowdown timescale but on the robust qualitative contrast between soliton-supported cores and cuspy NFW halos under identical assumptions, which consistently favors significantly weaker bar–halo coupling in the cored case.

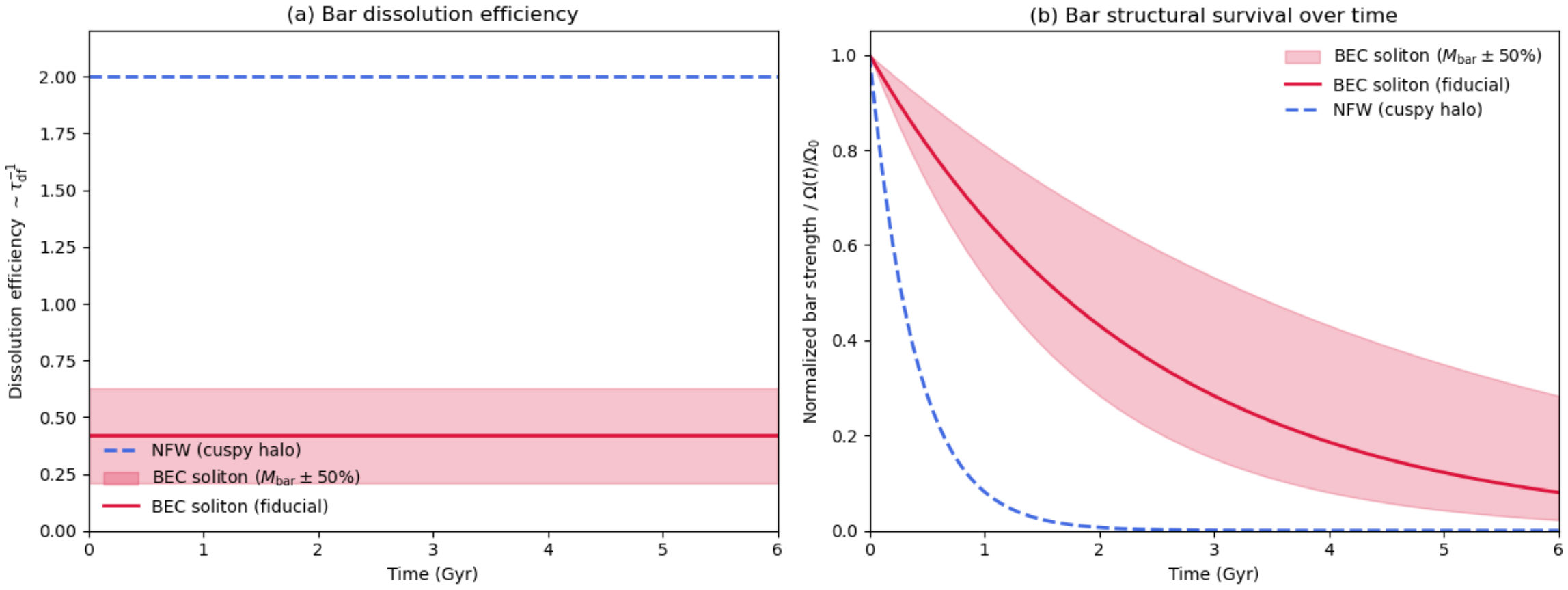

**Figure 4.** Bar dissolution efficiency (left) and normalized bar survival over time (right) for solitonic and NFW halos, illustrating the relative suppression of dynamical-friction-driven bar decay in a solitonic core.

Recent theoretical work incorporating wave-induced granularity in FDM further predicts an additional suppression of dynamical friction relative to the classical Chandrasekhar estimate (e.g., K. Blum et al. 2025). P. Boldrini et al. (2024) introduce a quantum correction factor $C_{\rm FDM}(kr,\ M_\sigma) < 1$, where $k = m_\chi v/\hbar$ is the wavenumber ($m_\chi \sim 10^{-22}$ eV, $v \sim 25$–$30\ {\rm km\ s^{-1}}$), $r$ is the orbital radius ($\sim$1 kpc), and $M_\sigma = v/\sigma_{\rm h}$ ($\sim$1–2). For DDO 168-like parameters ($kr \sim 2$–6, given a de Broglie wavelength $\lambda_{\rm dB} \sim 2$–4 kpc), $C_{\rm FDM}$ oscillates around 0.5–1, indicating a further suppression of dynamical friction relative to the classical Chandrasekhar estimate. Incorporating wave-induced suppression factors such as those predicted in FDM dynamical friction studies would further lengthen the slowdown timescales estimated here, strengthening rather than weakening the qualitative conclusion (L. Lancaster et al. 2020). This behavior is consistent with the role of the solitonic core in stabilizing the bar without altering our conservative, classical approach. More detailed FDM-specific treatments and full $N$-body or hydrodynamical simulations are expected to further refine these estimates.

We stress that the bar-stability analysis is intended as a qualitative consistency check rather than a quantitative prediction of bar pattern-speed evolution; the suppression of dynamical-friction-driven bar slowdown reflects the presence of a shallow central density profile and is therefore a generic feature of cored halos, with the solitonic core providing a physically motivated realization within the FDM framework.

## 5. Conclusion

We developed a solitonic dark matter profile based on the GPP formalism to model the rotation curve of DDO 168. Using updated SPARC rotation-curve data and a MCMC analysis, the inferred axion mass, of order $10^{-22}$ eV, corresponds to a solitonic core with $R_c \simeq 2.4$–2.6 kpc and central density $\rho_0 \simeq (3.5$–$4.0) \times 10^7 M_\odot\ {\rm kpc^{-3}}$, reproducing the observed inner rotation-curve behavior within the central few kiloparsecs. Residual deviations at larger radii indicate the likely presence of an additional diffuse halo component, suggesting future two-component soliton+NFW modeling.

Beyond the static mass distribution, we find that the solitonic core plays an important role in the dynamical evolution of the stellar bar. We find characteristic bar-slowdown timescales in the multigigayear regime for plausible halo dispersions ($\sigma_h \sim 15$–$30\ {\rm km\ s^{-1}}$) and Coulomb logarithms ($\ln\Lambda \sim 2$–5). These timescales are substantially longer than those expected in cuspy NFW halos, indicating that soliton-supported cores suppress bar–halo angular-momentum transfer and mitigate rapid bar dissolution. This behavior is qualitatively consistent with recent numerical simulations of bars embedded in cored or FDM halos, which show only modest (≈10%–30%) weakening over several gigayears, in contrast to rapid slowdown in cuspy halos.

Taken together, our results support a physical picture in which the quantum-pressure-supported solitonic core of DDO 168 not only governs the inner kinematic structure of the galaxy but also influences its secular evolution by allowing weak bars to persist over multigigayear timescales. This coupling between FDM cores and baryonic structures may therefore play an important role in shaping the longevity and morphology of bars in dwarf galaxies. Future work incorporating composite soliton+halo models and fully time-dependent dynamical simulations will be valuable to further test and refine these conclusions.

## Acknowledgments

We thank our colleagues and collaborators for helpful discussions and constructive feedback during the development of this work. We are especially grateful to Tiziana Di Matteo, our Scientific Editor, for her careful evaluation and guidance in improving the clarity and presentation of the manuscript, and we acknowledge the support of the Astrophysical Journal editorial and production team throughout the publication process. We also acknowledge the use of the SPARC database and the LITTLE-THINGS H I data. This work received no external funding.

## Appendix A
## Solitonic Core Fit to DDO 154 Using SPARC Data

In this appendix, we apply the same solitonic dark-matter framework used in the main text to the dwarf irregular galaxy

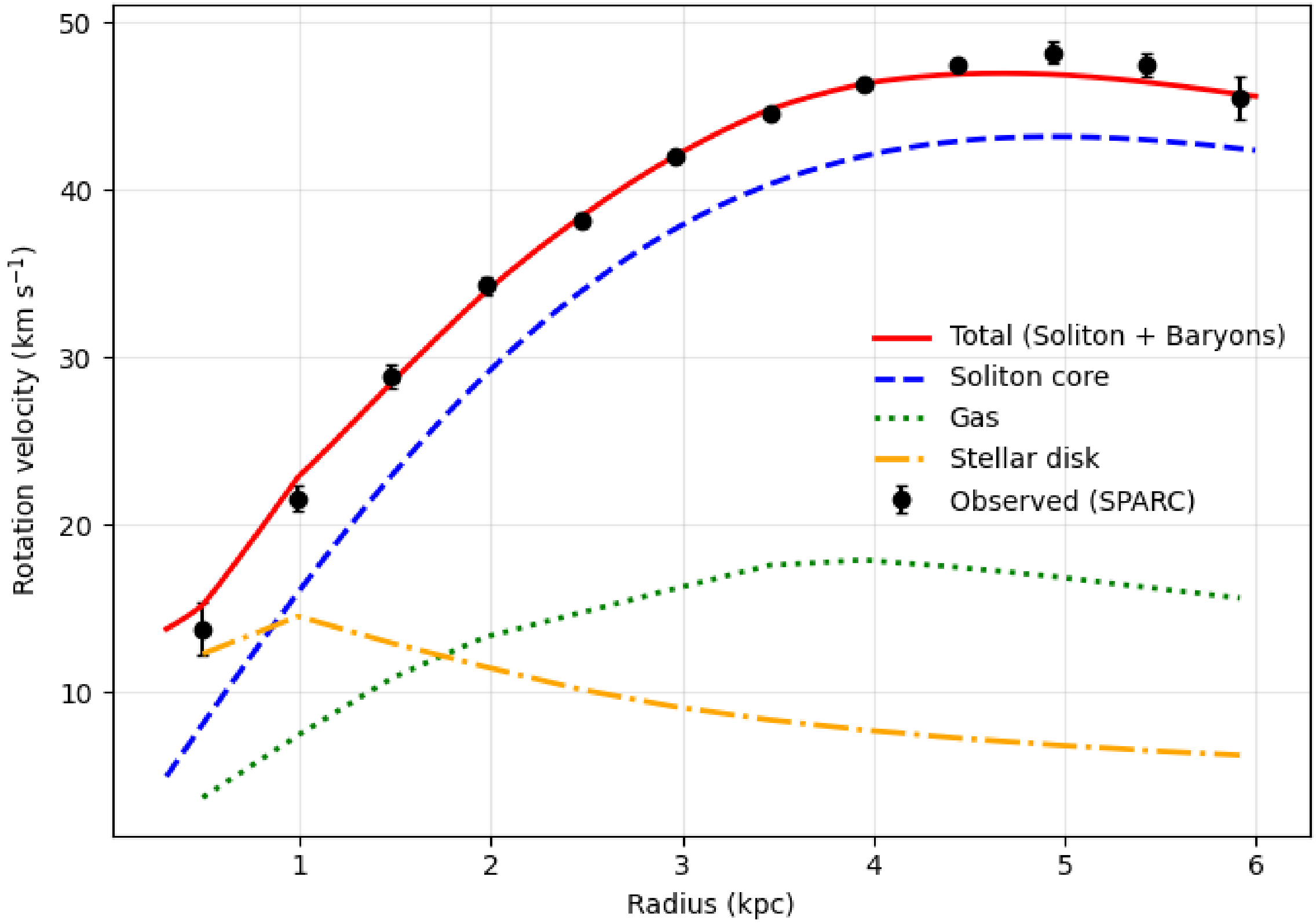

**Figure 5.** Rotation-curve decomposition of DDO 154 using SPARC data, showing the observed velocities with uncertainties and the best-fit solitonic core and baryonic contributions.

DDO 154, in order to provide an independent consistency check of the model. DDO 154 is a well-studied, dark-matter-dominated system with high-quality rotation-curve measurements, making it a suitable benchmark for testing soliton-based halo profiles.

We use the official SPARC rotation-curve data for DDO 154, including the published observational uncertainties, and decompose the observed circular velocity into contributions from gas, stellar disk, and a solitonic dark-matter core. The soliton density profile is modeled using the GPP solution, identical in functional form to that adopted for DDO 168 in the main text. No bar dynamics or secular evolution effects are considered in this appendix.

The total model rotation curve is constructed as

$$V_{\rm tot}^2(r) = V_{\rm soliton}^2(r) + V_{\rm gas}^2(r) + V_{\rm disk}^2(r), \tag{A1}$$

where the baryonic contributions are taken directly from the SPARC database and the solitonic core parameters are constrained through $\chi^2$ minimization with the soliton mass–radius scaling relation enforced, such that the central density is derived self-consistently from the two correlated soliton parameters, i.e., the axion mass $m$ and the core radius $R_c$.

Figure 5 shows the resulting best-fit solitonic core model for DDO 154, together with the individual baryonic components and the observed data points with error bars. The solitonic core provides an excellent description of the inner rotation curve, while the inclusion of baryons improves the fit at intermediate radii.

The best-fit soliton parameters for DDO 154 are:

1. Derived central density:

$$\rho_0 \simeq 1.55 \times 10^7 \, M_\odot \, {\rm kpc}^{-3}. \tag{A2}$$

2. Core radius:

$$R_c \simeq 2.53 \, {\rm kpc}. \tag{A3}$$

These values correspond to the minimum-$\chi^2$ solution; no statistical uncertainties are quoted here because this appendix is intended as a qualitative consistency check rather than a precision parameter inference.

The enclosed solitonic mass within $R = 2.96$ kpc is

$$M_{\rm soliton}(<2.96 \, {\rm kpc}) \simeq 9.75 \times 10^8 \, M_\odot. \tag{A4}$$

The enclosed solitonic mass is quoted at $R = 2.96$ kpc, corresponding to the outermost radius of the rotation-curve data used in the fit, rather than at the soliton core radius $R_c$, which characterizes the scale of the density profile but does not represent a physical truncation.

The resulting reduced chi-squared is

$$\chi_{\rm red}^2 \simeq 1.1, \tag{A5}$$

indicating a statistically acceptable fit given the observational uncertainties.

This appendix demonstrates that the solitonic core profile provides a consistent and observationally viable description of the inner rotation curve of DDO 154 when applied independently of the analysis presented for DDO 168. The results serve as an internal consistency check on the soliton modeling approach adopted in the main text, without invoking additional dynamical assumptions, and are not intended to establish universality across dwarf galaxies.

## Appendix B
## Bar–Halo Coupling in Cored and Cuspy Dark Matter Halos

We present here a controlled comparison of bar–halo dynamical interactions in two idealized halo models—solitonic (BEC/FDM) cores and cuspy NFW halos—using identical analytical methods and bar assumptions. Whereas the main text focuses on establishing the consistency of a solitonic core with the rotation-curve data of DDO 168, the purpose of this appendix is to examine how differences in the inner halo density profile alone affect the strength of bar–halo coupling, with all other inputs held fixed. The calculations are intended as a consistency check on the dynamical implications of cored versus cuspy halo structure rather than as a detailed model of bar evolution.

Figure 6 shows the resulting comparison for the dwarf galaxy DDO 154, used here as an independent benchmark system. The quantity plotted represents the radial dependence of the dynamical-friction efficiency associated with a stellar bar, evaluated using the Chandrasekhar formalism and expressed in normalized form. The solitonic core parameters for DDO 154 are adopted directly from Appendix A, while the NFW halo parameters are taken from published SPARC rotation-curve analyses (F. Lelli et al. 2016; H. Katz et al. 2017), without additional fitting. The bar mass and all other inputs are fixed to the same fiducial values and methodological assumptions used in the main text and applied consistently here.

The shaded regions indicate a conservative ±50% uncertainty in the assumed bar mass, propagated through the calculation. Over the radial range shown, the solitonic core exhibits a systematically lower dynamical-friction efficiency than the cuspy NFW halo. This separation persists across the full bar-mass uncertainty range, indicating that the difference arises from the underlying halo density structure rather than from the specific choice of bar parameters.

Because the efficiencies are shown in normalized form, the comparison is insensitive to the absolute normalization of the Chandrasekhar formula and should be interpreted as an order-of-magnitude comparison rather than a quantitative prediction of bar lifetimes. The qualitative behavior is similar to that obtained for DDO 168, suggesting that reduced bar–halo coupling is a generic feature of cored halo profiles within the assumptions and limitations of the present framework. The comparison presented here is intended as a qualitative, normalized diagnostic of bar–halo coupling rather than a quantitative prediction of bar lifetimes, and no claim is made regarding the detailed time evolution of bars in individual systems.

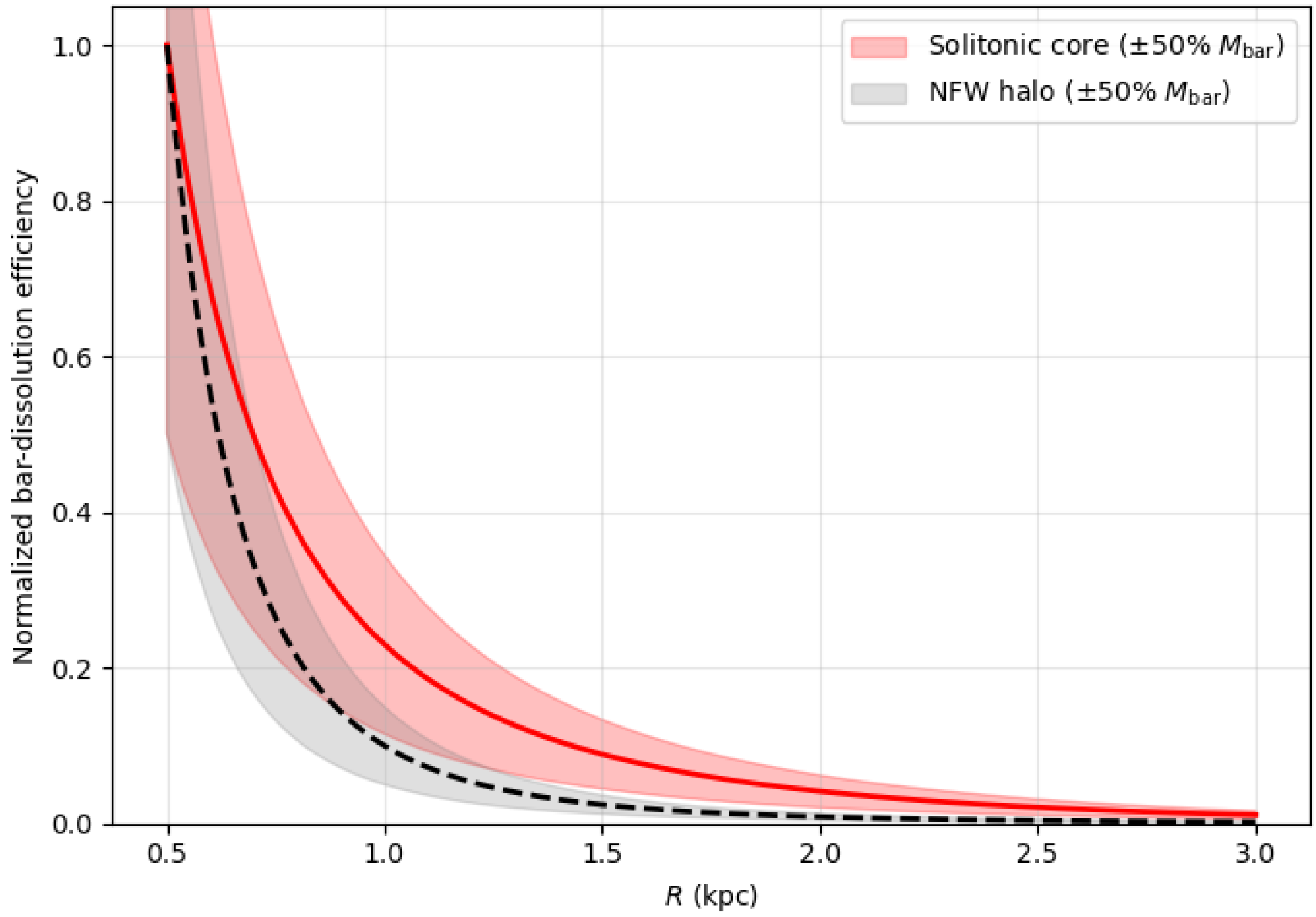

**Figure 6.** Normalized radial dependence of bar–halo dynamical-friction efficiency for DDO 154, comparing a solitonic core and a cuspy NFW halo, with shaded regions indicating a ±50% uncertainty in the assumed bar mass.

## ORCID iDs

Saroj Khanal https://orcid.org/0009-0001-3938-2806
Sanjay Kumar Sah https://orcid.org/0000-0002-8842-0442
Kiran Khanal https://orcid.org/0009-0000-0385-275X
Sapana Khanal https://orcid.org/0009-0008-6688-4909